\documentclass[%
 reprint,
 superscriptaddress,
 tightenlines,
aps,
pra,
 floatfix
]{revtex4-2}

\usepackage[american]{babel}
\usepackage{bm,graphicx,amsmath}
\usepackage{siunitx}
\usepackage{placeins}
\usepackage{amssymb}
\usepackage{amsmath}
\usepackage{epsfig}
\usepackage{nicefrac}
\usepackage{multirow}
\usepackage{mathtools}
\usepackage{graphicx}
\usepackage{braket} 
\usepackage[version=4]{mhchem}
\usepackage[colorlinks=true,linkcolor=blue,citecolor=blue,urlcolor=blue]{hyperref}
\usepackage{orcidlink}
\usepackage{tikz}
\usepackage{pgf}
\usepackage{pgfplots}
\usepackage{subfigure}
\pgfplotsset{compat=1.18} 
\usetikzlibrary{positioning, calc, matrix, arrows.meta, shapes.geometric, external, backgrounds}
\usepgfplotslibrary{external}
\DeclareSIUnit\sample{Sa}
\DeclareSIUnit\pixel{px}
\DeclareSIUnit\permille{\text{\textperthousand}}
\DeclareSIUnit{\belmilliwatt}{Bm}
\DeclareSIUnit{\dBm}{\deci\belmilliwatt}
    
\definecolor{cset-aps-limegreen}{RGB}{190,219,67}
\definecolor{cset-aps-green}{RGB}{31,138,112}

\newcommand*{\figref}[2][]{%
	\hyperref[{fig:#2}]{%
		Fig.~\ref*{fig:#2}%
		\ifx\\#1\\%
		\else
		(#1)%
		\fi
	}%
}

\newcommand*{\figureref}[2][]{%
	\hyperref[{fig:#2}]{%
		Figure~\ref*{fig:#2}%
		\ifx\\#1\\%
		\else
		(#1)%
		\fi
	}%
}

\newcommand*{\figuresref}[2][]{%
	\hyperref[{fig:#2}]{%
		Figures~\ref*{fig:#2}%
		\ifx\\#1\\%
		\else
		(#1)%
		\fi
	}%
}

\begin{document}
\title{Efficient Assembly of a Defect-Free Quantum Register of 1024 Neutral-Atom Qubits}

\author{Lukas Sturm\orcidlink{0009-0009-9130-0319}}
\affiliation{Technische Universit\"at Darmstadt, Institut f\"ur Angewandte Physik, Schlossgartenstra\ss e 7, 64289 Darmstadt, Germany}
\author{Marcel Mittenb\"uhler\orcidlink{0000-0002-1751-3158}}
\affiliation{Technische Universit\"at Darmstadt, Institut f\"ur Angewandte Physik, Schlossgartenstra\ss e 7, 64289 Darmstadt, Germany}
\author{Tim Gollerthan\orcidlink{0009-0005-5658-6017}}
\thanks{Present address: Universit\"at Innsbruck, Institut f\"ur Experimentalphysik, Technikerstra\ss e 25, 6020 Innsbruck, Austria}
\affiliation{Technische Universit\"at Darmstadt, Institut f\"ur Angewandte Physik, Schlossgartenstra\ss e 7, 64289 Darmstadt, Germany}
\author{Malte Schlosser\orcidlink{0000-0001-9004-5664}}
\affiliation{Technische Universit\"at Darmstadt, Institut f\"ur Angewandte Physik, Schlossgartenstra\ss e 7, 64289 Darmstadt, Germany}
\author{Gerhard Birkl\orcidlink{0000-0002-4137-9227}}
\email[For correspondence: ]{apqpub@physik.tu-darmstadt.de}
\homepage[\newline Homepage: ]{https://www.iap.tu-darmstadt.de/apq}
\affiliation{Technische Universit\"at Darmstadt, Institut f\"ur Angewandte Physik, Schlossgartenstra\ss e 7, 64289 Darmstadt, Germany}
\affiliation{Helmholtz Forschungsakademie Hessen f\"ur FAIR (HFHF), GSI Helmholtzzentrum für Schwerionenforschung, 64291 Darmstadt}
\date{\today}

\begin{abstract}
Low-entropy arrays of atomic quantum systems in optical tweezers offer unique prospects for fundamental research on few- and many-body systems as well as for extended applications in quantum technology.
The significance of this approach relies on the achievable system size, its uniformity, and the rate of qubit allocation. 
We propel the neutral-atom quantum-technology platform by the rapid assembly of a regular two-dimensional quantum register of up to 1024 atomic qubits, enabled by a novel implementation of intensity-homogenized tweezer arrays and parallelized atom transport.
Our highly efficient microoptical architecture modularizes intensity equalization and tweezer patterning in separate functional units, eliminating restrictions that arise for
high-power, high-resolution, and large-scale light-field control within a single device.
Arrays of precise grid structure, trap depth, and vibrational frequency with more than 3500 sites are demonstrated.
Individual sites are interconnected by up to 50 parallelized transport tweezers with intensity and position control in real-time for swift qubit relocation.
Multi-tweezer transport enables the operation of target patterns of up to 32$\times$32 sites with sustained near-unity filling fraction.
These results boost neutral-atom quantum information science above the kiloqubit level. 
\end{abstract}

\maketitle
\section{Introduction}
\label{sec:I}
Rooted in advanced theoretical methods and experimental capabilities that enable excellent control at the single-particle level, laser-optically generated arrays of ordered quantum systems represent one of the most sophisticated platforms for quantum science and technology \cite{Morgado2021,Gross2021,Amico2021,Grass2025}.
In tweezer arrays \cite{Dumke2002}, laser-cooled atoms and molecules are spatially positioned in a grid of focused laser beams in near perfect isolation with their quantum states being subject to comprehensive spatio-temporal control by electromagnetic fields \cite{Browaeys2020,Kaufman2021,Langen2024}. These arrays of naturally identical qubits with tunable interactions serve to put prime examples of long-range interacting quantum systems into practice \cite{Defenu2023}, targeting the frontiers of research in fundamental quantum effects and emerging applications \cite{Morgado2021,Wintersperger2023,Cornish2024,Shen2025,Halimeh2025}.

\begin{figure}[t]
	\includegraphics[width=\linewidth]{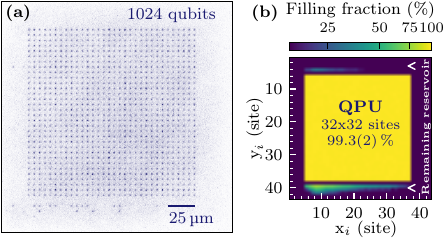}
	\caption{
		Large-scale array of atomic qubits forming a quantum processing unit (QPU) of {32$\times$32} sites and a surrounding reservoir in an array of {43$\times$43} homogenized optical tweezers created by modularized microoptical light field patterning. Inerconnectivity is enhanced by parallelized atomic qubit transport. (a) Single-shot fluorescence image of a defect-free QPU of \num{1024} qubits. (b) Site-resolved visualization of the cumulative maximum filling fraction for twelve successive rearrangement sequences.
        The QPU filling fraction is \SI{99.3(0.2)}{\percent}.}
    	\label{fig:First1024}
\end{figure}

Current architectures capitalize on a rich toolbox for controlling and shaping optical dipole traps that gives access to tweezer arrays generated by holographic spatial light modulation (SLM) \cite{Chew2024,Lin2025}, acousto-optic deflectors (AOD) \cite{Cooper2018,Yan2022,Radnaev2024, Mittenbuhler2025} or passive microstructured optical elements, such as microlens arrays (MLA) \cite{Dumke2002,Pause2024}, transmission masks \cite{Fang2025} and metasurfaces \cite{Hsu2022,Holman2026}.
Hybrid implementations \cite{Pause2023,Pause2024,Gyger2024,Norcia2024,Holzl2024,Nakamura2024,Grinkemeyer2025,Manetsch2025,Chiu2025,Li2025} use multiple technologies to combine static arrays, optical cavities and lattices, qubit reservoirs, and dynamic AOD-tweezers. This allowed pioneering demonstrations of single atom interfacing \cite{Shaw2025,Desantis2026}, quantum sensing and metrology \cite{Schaffner2024,Cao2024,Finkelstein2024,Kitson2025} as well as quantum simulation \cite{Qiao2025,Evered2025} and logical computation \cite{Reichardt2024,Chung2025,ZhangBichen2026,Bluvstein2026}.
In this article we present large-scale defect-free qubit arrays that propel our MLA approach above the one-kiloqubit level in a single plane (\figref{First1024}), enabled by a novel homogenized trap architecture and swift, parallelized, and repeated interconnectivity.

\begin{figure*}[t!]
    \centering
	\includegraphics[width=\linewidth]{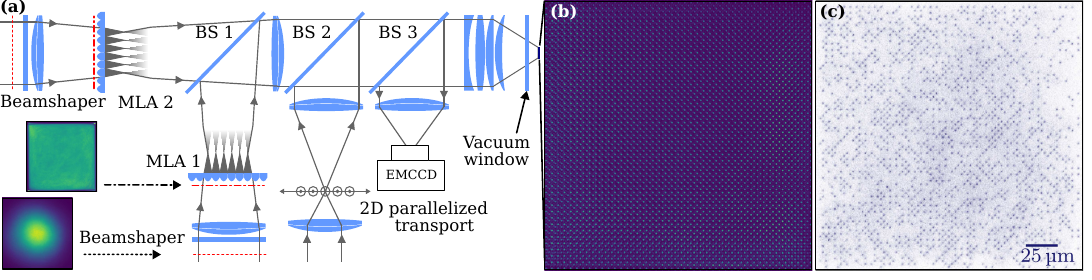}
	\caption{
(a) Schematic of the optical setup. In two independent input ports, a diffractive beamshaper converts an input Gaussian beam (dashed line) into a top-hat profile (dash-dotted line). Each beam illuminates a microlens array (MLA 1,2) generating a homogenized two-dimensional (2D) array of laser foci. The two resulting static tweezer arrays are combined using a polarizing beam splitter (BS 1) and reimaged into the vacuum chamber. Dynamically steerable 2D parallelized optical tweezers for atom transport are superimposed using a non-polarizing beam combiner (BS 2) with \SI{90}{\percent} transmission. (b) Intensity pattern of the combined MLA-generated trap arrays in an evenly-space interleaved configuration of more than \num{3500} sites. (c) Fluorescence image of 1707 trapped individual-atom qubits loaded from a magneto-optical trap and detected using an electron-multiplying charge-coupled device (EMCCD).
}\label{fig:SetupLightAtoms}
\end{figure*}
In general, neutral-atom-based implementations are characterized by a high degree of scalability due to the straightforward qubit supply \cite{Schlosser2012,Pause2023,Gyger2024,Norcia2024,Chiu2025,Li2025}, parallelized optical methods for state manipulation \cite{Schlosser2011,Graham2023,ZhangBichen2024,Bluvstein2026}, and trap array generation \cite{Schlosser2023,Pichard2024,Pause2024,Manetsch2025,Lin2025,Fang2025,Holman2026}.
However, the objective of implementing homogeneous systems of similar and sufficient trap depth leads to crucial technological challenges for scaling up the number of trapping sites and parallelizing atomic qubit transport operations.
While  MLAs excel at creating large-scale laser spot patterns \cite{Schlosser2011,Schlosser2023,Pause2024} and maintain uniform optical efficiency when increasing the number of sites, also in this platform, a significant challenge arises from the direct mapping of the Gaussian input laser beam profile onto the intensity distribution of the generated tweezer arrays. 
Furthermore, interconnectivity in the static grid of MLA-tweezers has to be conveyed by high-bandwidth AOD-steerable atomic qubit transport \cite{Barredo2016,Endres2016}.
Here, favorable parallelized multi-tweezer operation is prone to uncontrolled trap modulation during dynamic trajectories originating from the nonlinear acoustooptic response and intermodulation \cite{Endres2016,Tian2023}. 
This has led to the application of non-scalable mitigation methods based on precalibration of a fixed set of patterns or trajectories and limited the usable number of parallelized dynamic tweezers \cite{Tian2023}, restraining transport rates and conditional flexibility.
We resolve these challenges for efficient and sustainable qubit-array allocation by combining scalable, high-power tweezer-array generation via passive MLAs with passive optical beam shaping for homogenization and precise real-time intensity control in dynamic multi-tweezer beam steering \cite{Mittenbuhler2025}.\\

%%%%%%%%%%%%%%%%%%%%%%%%%%%%%%%%%%%%%%%%%%%%%%%%%%%%%%%%%%%%%%%%%%%%
\section{Microoptical Generation of Homogenized Tweezer Arrays}
\label{sec:II}
%%%%%%%%%%%%%%%%%%%%%%%%%%%%%%%%%%%%%%%%%%%%%%%%%%%%%%%%%%%%%%%%%%%%%
Our platform separates MLA-based generation of a large-scale tweezer array for atom trapping from spatial shaping of a uniform intensity profile for MLA illumination. This modularization allows for independent optimization of both functions.
The resulting homogenized trap array facilitates equalized loading rates \cite{Schymik2022} and significantly improved transport fidelities. Both pioneer the first application of extensively parallelized dynamic atom rearrangement \cite{Mittenbuhler2025} for tweezer-array supercharging \cite{Pause2024} and fast large-scale assembly of defect-free arrays of individual atoms.

\subsection{Experimental Setup}
A schematic of the experimental setup is presented in \figref[a]{SetupLightAtoms}. Two interleaved MLA-based quadratic-grid tweezer arrays of \num{43}$\times$\num{43} sites form a two-dimensional (2D) array of more than \num{3500} traps.
Each MLA is illuminated by a trapping laser beam with a homogenized top-hat profile that is converted from a Gaussian input beam by a diffractive optical beamshaper with $\SI{94.5}{\percent}$ efficiency. 

The focal planes of the MLAs are superimposed by a polarizing beamsplitter (BS 1)
and reimaged into a vacuum chamber forming two interleaved regular arrays of dipole traps with \SI{0.86(7)}{\micro\meter} waist and \SI{5.16(3)}{\micro\meter} pitch.
Combined in a symmetric configuration, this results in more than \num{3500} trap sites with \SI{3.65(3)}{\micro\meter} spacing (\figref[b]{SetupLightAtoms}). The modular architecture circumvents laser power limitations as it renders the independent creation of multiple homogenized tweezer arrays and their scalable combination with high overall efficiency possible. 
Individual laser-cooled rubidium atoms serving as single-atom qubits are trapped in both arrays and detected by fluorescence imaging (\figref[c]{SetupLightAtoms}) with average fidelity of \SI{99.8(2)}{\percent} \cite{Pause2024}. The combined array can be used as quantum memory or processor with the full site count. In the work presented here, we designate one array as processing array, whereas the other array serves as a qubit reservoir for initial high-efficiency loading. Atoms from the reservoir array are used to supercharge the processor array \cite{Pause2024} which embeds the freely selectable target pattern intended as quantum processing unit (QPU). Supercharging significantly increases the number of atomic qubits that are available for QPU assembly.

A non-polarizing beam combiner (BS 2) is used for superimposing the two static trap arrays with a steerable 2D tweezer array for atom transport and target pattern assembly.
The dynamic system employs two perpendicular AODs for beam steering along both array dimensions. The transport tweezer waist is \SI{1.02(2)}{\micro\meter} while the addressable area covers {50$\times$50} sites surpassing the spatial extent of the static arrays.
Real-time control for strict intensity matching of multi-frequency trajectories \cite{Mittenbuhler2025} enables the low-latency execution of random multi-tweezer transport operations as long as they can be parallelized along a straight line in the 2D array.

\subsection{Tweezer-Array Homogeneity}
\begin{figure}
	\includegraphics[width=\linewidth]{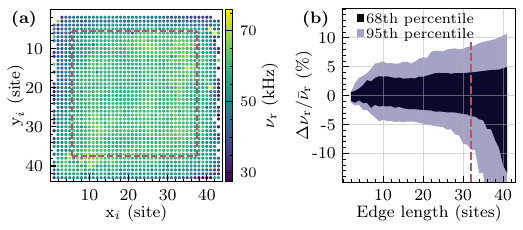}
	\caption{
    Radial vibrational frequency $\nu_\text{r}$ in the homogenized {43$\times$43} site static tweezer array with the QPU ({32$\times$32} sites) indicated by dashed lines.
    (a) Site-resolved, color-coded representation of $\nu_\text{r}$. Sites with insufficient statistics are left blank. 
    (b) Spread of $\nu_\text{r}$ as deviation from the median for increasing edge length of a central quadratic array. The 68th (black) and 95th (blue) percentiles are shown.}
    \label{fig:TrapFrequencyUniformity}
\end{figure}
We characterize the uniformity of the MLA-based processor array homogenized by the top-hat converter by measuring the radial vibrational frequency $\nu_\text{r}$ in each trap. For a trap depth that significantly exceeds the thermal energy of the atoms, $\nu_\text{r}$ represents the most important property of each trap comprising variations in both the trap depth $U$ and the trap waist $w_0$ ($\nu_\text{r}\propto\sqrt{U/w_0^2}$). Using vibrational excitation by successively switching off and on the traps twice with varying intermediate delay, the modulated atom recapture efficiency after the final turn-on reveals the vibrational frequencies. 

The site-resolved data are displayed for a homogenized array of 43$\times$43 sites in \figref[a]{TrapFrequencyUniformity}. Sites with insufficient statistics are left blank. 
We can operate a homogeneous central region of 41$\times$41 sites. For the targeted QPU size of {32$\times$32} sites (dashed line), the median trap frequency with 68th percentile deviations is $\tilde{\nu}_\text{r}=\SI{60.4(2.4:2.2)}{\kilo\hertz}$. This region shows a rms uniformity, given as the ratio of the rms deviation $\Delta \nu_\text{r}$ and the mean $\bar{\nu}_\text{r}$, of $\Delta \nu_\text{r} / \bar{\nu}_\text{r} = \SI{4.5}{\percent}$. With the measured trap frequency and waist, we calculate a trap depth of $U=k_B\cdot\SI{273(75)}{\micro\kelvin}$. In an independent measurement, we have determined the temperature of the atoms in the traps to \SI{17(2)}{\micro\kelvin}.

\figureref[b]{TrapFrequencyUniformity} presents an analysis of the variation of the vibrational frequency as a function of the edge length of a quadratic trap array. Displayed are the 68th (black) and 95th (blue) percentiles of the deviation from median.

We attribute the increased spread of frequencies for large arrays to edge effects of the top-hat beam conversion that cause a modulation of the intensity distribution and the $k$-vector spectrum for sites close to the edge of the top-hat beam. 
The observed spread is in accordance with the variation of the intensity distribution measured for the respective region of the top-hat beam illuminating the MLA.
The uniformity and the total number of traps significantly exceed the values achievable by illuminating the MLA by a Gaussian beam with the same effective laser power, i.e., incorporting the efficiency of the top-hat conversion. A Gaussian beam size optimizing the number of traps with a depth equal or deeper than the homogenized array average results in 667 traps with the Gaussian intensity envelope deteriorating the uniformity to $\SI{21.1}{\percent}$ for 667 sites in contrast to the \SI{4.5}{\percent} uniformity of the 1024-site array presented here.

%%%%%%%%%%%%%%%%%%%%%%%%%%%%%%%%%%%%%%%%%%%%%%%%%%%%%%%%%%%%%%%%%%%%%
\section{Qubit Pattern Assembly}
\label{sec:III}
%%%%%%%%%%%%%%%%%%%%%%%%%%%%%%%%%%%%%%%%%%%%%%%%%%%%%%%%%%%%%%%%%%%%%
\begin{figure*}[t!]
    \centering 
	\includegraphics[width=\linewidth]{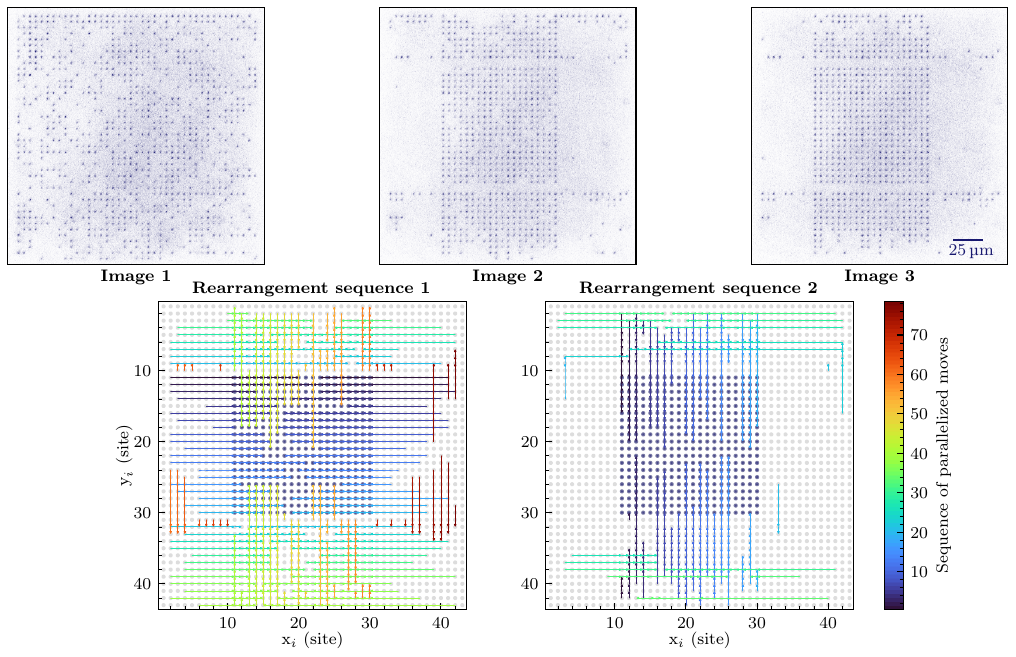}
	\caption{
Parallelized rearrangement of atomic qubits into a defect-free 20$\times$20 target pattern within two rearrangement sequences. Three subsequent fluorescence images (top) and the two interposed rearrangement sequences (bottom) containing consecutive multi-tweezer transport trajectories are shown. The color code indicates the temporal order of parallelized moves along rows and columns within each rearrangement sequence. Prior to the image 1, atom loading of both static arrays and supercharging of the processor array are performed.}
	\label{fig:moves}
\end{figure*}

\subsection{Parallelized Multi-Tweezer Atom Rearrangement} \label{sec:rearrangement}
Defect-free qubit arrays are created by initial supercharging the processor array out of the reservoir array and subsequent rearranging the atoms of an imperfect initial configuration into a defect-free target pattern within the processor array using a dynamic AOD-based 2D transport-tweezer array. 
The transport tweezers are calibrated to meet the trap positions of the static arrays for optimal atom pick-up and delivery. A rearrangement sequence comprises a $t_0 = \SI{78.6(1)}{\milli\second}$ phase of atom detection (\SI{50}{\milli\second}), image readout, data processing and experimental waits, followed by successive one-dimensional (1D) transport operations carried out for up to \num{43} tweezers in parallel along rows and columns. Up to \num{50} repetitions of rearrangement sequences are applied to increase the cumulative success rate for obtaining a defect-free target pattern.

Atoms are reloaded between tweezers and transported by applying intensity ramps and beam steering in sequence. Within the initial and the final \SI{180}{\micro\second} the transport tweezers' vibrational frequency is linearly ramped between \num{0} and twice the median frequency of the static array for atom pick-up and vice-versa for delivery. For atom transport, a constant jerk trajectory is implemented \cite{Manetsch2025}. The jerk is defined as the derivative of the acceleration. 
We set the jerk to \SI{1.09e5}{\micro\meter\per\cubic\milli\second} and limit the maximum acceleration to \SI{5.45e3}{\micro\meter\per\square\milli\second} and the maximum velocity to \SI{81.8}{\micro\meter\per\milli\second}. The temporal profile of acceleration is reversed for deceleration during the second half of the transport trajectory.
A single-pitch transport requires $\SI{478}{\micro\second}$ in total with \SI{128}{\micro\second} of atom motion. In the case of parallelized transport including a range of distances, all trajectories are modified to match the duration of the longest path.
The algorithm proceeds through deterministic, alternating passes that reduce the 2D problem into successive 1D matching tasks. For each 1D pass, a greedy nearest-assignment algorithm maps available atoms to empty target sites by minimizing transport distance while preserving order, i.e., collision-free parallel motion and compression moves.
Initial supercharging between arrays is implemented by matching reservoir-array atoms with empty nearest neighbors of the processor array. Parallelized transport is executed consecutively for all four spatial directions.

\figureref{moves} illustrates the assembly of a defect-free 20$\times$20 qubit target pattern after supercharging of the processor array. The top row shows three consecutive fluorescence images of the atom array, while the bottom row depicts the two interposed rearrangement sequences. The color code indicates the temporal order of parallelized moves along rows and columns within each rearrangement sequence. If any column intersecting the target pattern does not contain the minimum number of atoms required, row-wise operations rearrange atoms first before refining the target pattern column-wise. This is the case for rearrangement sequence 1. Otherwise, column-based target sorting is performed directly, as in rearrangement sequence 2 that attains the defect-free pattern. At the end of each rearrangement sequence surplus atoms are organized into structured bands above, below, and on both sides of the target pattern. These reservoirs facilitate for subsequent rearrangement sequences to compensate for remaining and emerging defects resulting from non-perfect transport as well as atom loss arising due to finite trap lifetime or loss-based detection. Our approach combines dimensional reduction with reservoir-aware heuristics to enable efficient and reproducible rearrangement of neutral atom arrays. The computational complexity scales as ($O(N^2)$) per line, resulting in a fully deterministic and experimentally tractable rearrangement sequence suitable for large-scale atom assembly.

\subsection{Efficient Assembly of Dense Qubit Arrays}
\begin{figure}[t]
\includegraphics[width=1\linewidth]{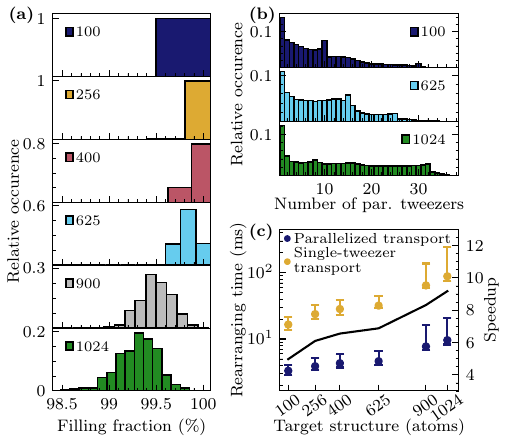}
\caption{Characteristics of parallelized assembly of target patterns up to 1024 qubits in a dense quadratic geometry.
(a) Cumulative maximum filling statistics for the assembly of target patterns with \numrange{100}{1024} atomic qubits using multiple rearrangement sequences.
(b) Multi-tweezer parallelization: Relative occurrence of the number of simultaneously applied transport tweezers in the assembly of target patterns with 100, 625, and 1024 sites.
(c) Rearrangement time $\tilde{t}_\text{r,(M,S)}$ for sequential single-tweezer transport (yellow) and parallelized transport (blue) given as the median of the combined duration of all transport operations within one rearrangement sequence. By parallization, a speedup (line) of a factor of 9 is achieved for the \num{1024} atom target structure.}
\label{fig:Speedup}
\end{figure}
We characterize the efficiency of the introduced methods by implementing, analyzing, and modeling the parallelized assembly of target patterns up to 1024 atomic qubits in a dense quadratic geometry. On average \SI{1292(30)} atoms are available in the processor array after supercharging, allowing multiple repetitions of rearrangement sequences. \figureref[a]{Speedup} gives the statistics of the maximum filling that has been achieved as a function of the target-pattern size. The data were obtained from 1000 experimental runs each. For patterns up to \num{625} sites, we set the number of repetitions to $n_\text{r}=\num{50}$, while the remaining reservoir atoms would have sufficed for even higher $n_\text{r}$. For larger target patterns, a depletion of the reservoir for late rearrangement sequences causes a decline in the average filling fraction which imposes a expedient limit of $n_\text{r}=\num{15}$ for \num{900} and $n_\text{r}=\num{12}$ for \num{1024} sites.

Target patterns up to \num{256} qubits always reach complete filling with a slight degression of \SI{100.00(0:25)}{\percent} for the cumulative maximum filling fraction for 400 qubits.
For larger QPUs, the role of defects emerging during a rearrangement sequence becomes more pronounced, limiting the cumulative maximum filling to \{\SI{99.84(16:16)}{\percent}, \SI{99.44(22:11)}{\percent}, \SI{99.32(19:20)}{\percent}\} for a number of $n_q = \{625,900,1024\}$ sites.

To quantify the importance of parallelization of the transport operations, an analysis of all rearrangement sequences underlying the data set of \figref[a]{Speedup} has been conducted.
\figureref[b]{Speedup} displays the relative occurrence of the number of tweezers used in parallelized multi-tweezer transports for the three target pattern sizes of $n_q = \{100,~625,~1024\}$ qubits, giving a median number 
of parallelized tweezers of \{\SI{7(9:5)},~\SI{10(7:8)},~\SI{13(14:11)}\}, respectively. 
While the number of parallelized tweezers predominantly lies within the linear size of the target patterns of $\sqrt{n_q}=\{10,~25,~32\}$ sites, observable as a drop in the relative occurence at these tweezer counts in \figref[b]{Speedup}, all sites of the array potentially supply the target pattern and the surrounding surplus areas. This manifests in the occurrence of transport operations with up to 38 tweezers for all target pattern sizes.

The survival probability of a transported atom averaged over all multi-tweezer configurations is ${\bar{\eta}_\text{exp}=\SI{98.5(3)}{\percent}}$ per rearrangements sequence after the first three sequences. During the first three sequences a reduced value of $\bar{\eta}_\text{exp}$ is observed.
We assign this reduction to nonthermalized atoms only weakly trapped directly after tweezer loading, increased collisional losses caused by a remaining local background of laser cooled atoms shortly after the MOT phase, prolonged rearrangement sequences due to a high number of moves required during initial construction of the target pattern and reservoir area, and technical effects such as delayed thermalization of the AODs.
For further analysis, we embrace all statistical loss mechanisms in an average effective lifetime $\tau_\text{eff}$ and describe the survival probability after time $t$ as \begin{align}\label{eqn:peff}
\bar{p}=\exp{(-t/\tau_\text{eff}).}
\end{align}

This gives a good approximation for the combined effect of all loss processes, in which loss rates depend on a multitude of parameters, potentially being subject to changes at different experimental stages, such as imaging and rearrangement, as well.
A value for losses due to background gas collisions, atom heating, and imaging-induced losses was determined in a separate measurement to $\tau_\text{static}=\SI{10(1)}{\second}$ for atoms at rest in static MLA-based arrays. This leads to a lower bound of \SI{0.8(1)}{\percent} loss for $t = t_0$, in itself being the lower bound for the duration of a rearrangement sequence.
Together with ${\bar{\eta}_\text{exp}=\SI{98.5(3)}{\percent}}$, the estimated transport efficiency is at least \SI{99.3(3)}{\percent}.

The evident vacuum limit substantiates the imperative of swift transport for the efficient creation of large-scale, defect-free neutral atom QPUs, as pursued by the parallelization of transport presented in this work. \figureref[c]{Speedup} compares the experimentally observed multi-tweezer transport times to the calculated expenditure for carrying out all moves in a sequential single-tweezer approach. The data display the median duration $\tilde{t}_\text{r,M}$ of all recorded rearrangements for the given target sizes with parallelized transport. Parallelized sequences outperform single-tweezer sequences by speedup factors of \numrange{5}{9}. For \num{1024} qubits, we achieve millisecond-regime parallelized rearrangement durations with a median of $\tilde{t}_\text{r,M}=\SI{9.6}{\milli\second}$ as compared to $\tilde{t}_\text{r,S}=\SI{88.1}{\milli\second}$ for single-tweezer transport.

The implications for the filling fraction after a single experimental rearrangement sequence are described by inserting the median duration of a rearrangement sequence $t_\text{(M,S)}=t_0+\tilde{t}_\text{r,(M,S)}$ as time $t$ in \autoref{eqn:peff} and computing $\bar{p}_\text{(M,S)}$. 
The average survival probability $\bar{p}_\text{(M,S)}$ is equivalent to the expected filling, independent of the particular rearrangement trajectory for a specific atom, as $\tau_\text{eff}$ incorporates the loss rate averaged for all atoms, whether transported or not.
The execution of each rearrangement sequence represents a Bernoulli trial with survival probability $\bar{p}_\text{(M,S)}$ for each atomic qubit at $n_\text{q}$ sites, described by the binomial probability mass function $f(k,n_\text{q},\bar{p}_\text{(M,S)})$. Accordingly, $k$ successes imply a filling fraction of $k/n_\text{q}$ and the mean filling is $\bar{p}_\text{(M,S)}$.
In the case of vanishing transport-induced losses, $\bar{p}_\text{(M,S)}$ approaches the expectation value for the remaining filling fraction after a given wait time $t_\text{(M,S)}$ with ${\tau_\text{eff}\rightarrow\tau_\text{static}}$.
With enough extra atoms in the reservoir sections outside the target QPU, 
the repeated application of $n_\text{r}$ rearrangement sequences is possible.
This significantly enhances the probability of achieving at least one defect-free target pattern during any experimental run. This can be quantified by analyzing the cumulative maximum filling fraction $\bar{p}_\text{cum}$ given as the average of the filling fractions in \figref[a]{Speedup} in the following effective model:
When regarding $\bar{p}_\text{cum}$ as statistical outcome of an effective \textit{single} rearrangement sequence with probability $\bar{p}_\text{cum}$ sufficing \autoref{eqn:peff}, cumulative repetitions that increase $\bar{p}_\text{cum}$ counteract lifetime-induced losses. This can be parametrized by a modified effective lifetime ${\tau_\text{eff}}$, potentially surpassing $\tau_\text{static}$ since it has the same effect as reduced losses due to improved vacuum conditions.

For parallized transport, the expectation value of the corresponding probability mass function of $\bar{p}_\text{cum}$ scales as
\begin{align}\label{eqn:pbb}
\bar{p}_\text{cum}=\frac{1}{n_\text{q}}\sum_{k=0}^{n_\text{q}-1} \{1-[F(k;n_\text{q},\bar{p}_\text{M})]^{n_\text{r}}\}
\end{align}\\
This expression connects the cumulative filling fraction $\bar{p}_\text{cum}$ with the binomial cumulative distribution function $F(k;n_\text{q},\bar{p}_\text{M})$ of a single experimental trial given by a single multi-tweezer rearrangement sequence with expected filling $\bar{p}_\text{M}$.
\begin{figure}[t]
	\includegraphics[width=1\linewidth]{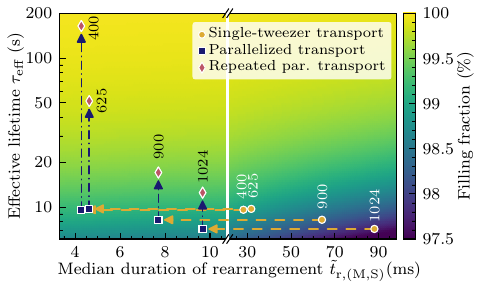}
	\caption{
    Color-coded representation of the mean cumulative maximum filling fraction for the assembly of $n_q = \{400,625,900,1024\}$ site target patterns using single-repetition single-tweezer transport (yellow circles), single-repetition parallelized transport (blue squares), and $n_\text{r}$ repetitions of parallelized transport (red diamonds). The experimental data for repetitive parallelized transport are used to infer filling fractions for the other two transport protocols (see main text). The median duration of rearrangements is given along the horizontal axis. Calculated effective lifetimes 
    $\tau_\text{eff}$ are plotted along the vertical axis. The reduction in rearrangement duration by parallelized transport (yellow arrows) and the application of $n_\text{r}$ repetitions increase the filling fraction. Repetitive transport raises the effective lifetime well beyond the measured vacuum lifetime $\tau_\text{static}$ (blue arrows).  }
	\label{fig:effVacVStiming}
\end{figure}

\figureref{effVacVStiming} gives a color-coded density plot of the maximum filling fractions for the assembly of $n_\text{q}=\{400, 625, 900, 1024\}$ site target patterns using single-repetition single-tweezer transport (yellow circles), single-repetition parallelized transport (blue squares), and $n_\text{r}$ repetitions of parallelized transport (red diamonds). The experimental data for repetitive parallelized transport of \figref[a]{Speedup} are used to calculate the expected values for the other two transport protocols.
The median rearrangement duration is given along the horizontal axis. 
Yellow arrows indicate the speedup achieved by replacing single-tweezer by parallelized multi-tweezer transport and blue arrows the gain in filling fraction by implementing $n_\text{r}$ repetitions.
The median duration of a rearrangement sequence results in $t_\text{(M,S)}=t_0+\tilde{t}_\text{r,(M,S)}$.
While $t_0$ remains constant, $\tilde{t}_\text{r,(M,S)}$ vary for different $n_\text{q}$ and transport protocols according to \figref[c]{Speedup}. This separates the results for single-tweezer and multi-tweezer protocols into two temporal domains in \figref{effVacVStiming}.

Using the total duration $t_\text{(M,S)}$, the number of repetitions $n_\text{r}$, and $\bar{p}_\text{cum}$ as inputs, the corresponding values of $\tau_\text{eff}$ can be calculated from \autoref{eqn:peff}. Probabilities $\bar{p}_\text{M}$ are obtained by numerically solving \autoref{eqn:pbb}. Subsequently, values for $\bar{p}_\text{S}$ are calculated for $t_\text{S}$ assuming the same $\tau_\text{eff}$ as for single-repetition parallelized transport.
For $n_\text{q}=400(1024)$ site target patterns, parallelizing transport reduces the defect probability $1-\bar{p}$ by a factor of 1.3(1.9). After the execution of $n_\text{r}$ repetitions, the reduction of the cumulative defect probability is 22.8(3.2), where the gain for large target patterns is only limited by the required reduction of $n_\text{r}$ due to depletion of reservoir atoms.

%%%%%%%%%%%%%%%%%%%%%%%%%%%%%%%%%%%%%%%%%%%%%%%%%%%%%%%%%%%%%%%%%%%%%
\section{Conclusion}
\label{sec:V}
%%%%%%%%%%%%%%%%%%%%%%%%%%%%%%%%%%%%%%%%%%%%%%%%%%%%%%%%%%%%%%%%%%%%%
In conclusion, we demonstrated the rapid assembly of large-scale, regular, two-dimensional arrays of neutral-atom qubits in homogenized optical tweezer arrays with more than 3500 sites, achieving a sustained near-unity filling fraction of \SI{99.3(2)}{\percent} for a 1024-site QPU. Our approach overcomes the conventional scaling limitations by modularizing intensity equalization and tweezer patterning in an efficient microoptical platform that holds unique potential for further scaling with MLAs that create millions of tweezers. The implementation based on passive optics circumvents limitations in laser power, damage thresholds, and finite diffraction efficiencies occurring for active spatial light patterning. 
We observe no perceptible degradation of the MLA-generated tweezer array when implementing the homogenized intensity profile. This architecture realizes an excellent trap-frequency uniformity of \SI{4.5}{\percent} within the central $32\times32$ sites by purely passive means. This can be further improved by a dedicated redesign of the beam shaper for MLA applications or by addition of active control elements as demonstrated in \cite{Schaffner2020}. The achievable system size can easily be scaled by modified beam shaping parameters and increased laser power.
As second important advance, we demonstrated highly parallelized, multi-tweezer atom transport that utilizes up to 38 dynamic tweezers simultaneously.  Enabled by real-time intensity control of multi-tweezer trajectories in combination with the homogenized trap array, the transport efficiency consistently reaches at least \SI{99.3(3)}{\percent}. This extensive, high-fidelity parallelization yields a speedup of a factor of 9 compared to sequential single-tweezer moves, reducing the median rearrangement time for a one kiloqubit QPU to \SI{9.6}{\milli\second}.

For future scaling, clear optimization routes are defined by the counterplay of rearrangement duration and transport efficiency on one side versus lifetime-induced losses on the other side, for which reduced imaging overhead and improved vacuum conditions are central.
%will unlock the full ability of the multi-tweezer speedup. 
An alternative approach for mitigating lifetime limitations was demonstrated in this work by implementing repetitive target-structure assembly. This technique can be described as an effective increase of the atom lifetime and can be applied whenever enough spare atoms are available. A combination of this technique with factually improved vacuum conditions and optimized transport trajectories will be the optimal approach:
For realistically accessible vacuum-limited lifetimes on the scale of \num{100} seconds, single-repetition filling of a {32$\times$32}-site target pattern will quickly approach a filling fraction of $\bar{p}=\num{0.999}$ after a single rearrangement sequence (calculated for $\tau_\text{eff}=\SI{135}{\second}$) limited by current state-of-the-art transport efficiency of \SI{99.9}{\percent} \cite{Manetsch2025,Chiu2025}. In case of dominating transport-induced losses, the above introduced model shows that only one additional repetition could largely compensate for remaining defects and guarantee unity filling.
This again underlines the impact of transport parallelization and trap homogenization for static and dynamic tweezers presented in this work, which enables efficient atomic qubit rearrangement paving the way for next-generation quantum information science and technology including utility-scale quantum computation \cite{Cain2026,Zhao2026,Menssen2026}.

\begin{acknowledgments}
We acknowledge financial support by the Federal Ministry of Education and Research (BMBF) [Grant No. 13N15981], by the Federal Ministry of Research, Technology and Space (BMFTR) [Grant No. 13N17366 and 13N17521], and by the Deutsche Forschungsgemeinschaft (DFG -- German Research Foundation) [Grant No. BI 647/6-1 and BI 647/6-2, Priority Program SPP 1929 (GiRyd)]. During the completion of this manuscript, we became aware of a complementary approach for increasing assembly efficiency by improved vacuum conditions \cite{Lim2026}.
\end{acknowledgments}

\section*{DATA AVAILABILITY}
The data that support the findings of this article are not publicly available because they contain commercially sensitive information. The data are available from the authors upon reasonable request.

\bibliography{Parallel}

@article{Amico2021,
	author = {Amico, L and Boshier, M and Birkl, G and Minguzzi, Anna and Miniatura, C and Kwek, L-C and Aghamalyan, D and Ahufinger, V and Anderson, D and Andrei, Natan and others},
	title = {Roadmap on Atomtronics: State of the art and perspective},
	journal = {AVS Quantum Science},
	volume = {3},
	number = {3},
	pages = {039201},
	year = {2021},
	doi = {10.1116/5.0026178},
	URL = {https://doi.org/10.1116/5.0026178}
}

@article{Barredo2016,
	author = {Barredo, Daniel and de L{\'e}s{\'e}leuc, Sylvain and Lienhard, Vincent and Lahaye, Thierry and Browaeys, Antoine},
	title = {An atom-by-atom assembler of defect-free arbitrary two-dimensional atomic arrays},
	volume = {354},
	number = {6315},
	pages = {1021--1023},
	year = {2016},
	doi = {10.1126/science.aah3778},
	publisher = {American Association for the Advancement of Science},
	issn = {0036-8075},
	URL = {http://science.sciencemag.org/content/354/6315/1021},
	journal = {Science}
}

@article{Chung2025,
  title   = {Fault-tolerant operation and materials science with neutral atom logical qubits},
  author  = {Chung, Woo Chang and Cole, Daniel C. and Gokhale, Pranav and Jones, Eric B. and Kuper, Kevin W. and Mason, David and Omole, Victory and Radnaev, Alexander G. and Rines, Rich and Teo, Mariesa H. and Bedalov, Matt J. and Blakely, Matt and Buttler, Peter D. and Carnahan, Caitlin and Chong, Frederic T. and Goiporia, Palash and Heim, Bettina and Hickman, Garrett T. and Jones, Ryan A. and Khalate, Pradnya and Kim, Jin-Sung and Lichtman, Martin T. and Lee, Stephanie and Neff-Mallon, Nathan A. and Noel, Thomas W. and Saffman, Mark and Shabtai, Efrat and Thotakura, Bharath and Tomesh, Teague and Tucker, Angela K.},
  journal = {npj Quantum Information},
  volume  = {11},
  pages   = {193},
  year    = {2025},
  doi     = {10.1038/s41534-025-01095-w},
  url     = {https://doi.org/10.1038/s41534-025-01095-w},
	publisher={Nature Publishing Group}
}

@article{Bluvstein2026,
  title={A fault-tolerant neutral-atom architecture for universal quantum computation},
  author={Bluvstein, Dolev and Geim, Alexandra A. and Li, Sophie H. and Evered, Simon J. and Bonilla Ataides, J. Pablo and Baranes, Gefen and Gu, Andi and Manovitz, Tom and Xu, Muqing and Kalinowski, Marcin and Majidy, Shayan and Kokail, Christian and Maskara, Nishad and Trapp, Elias C. and Stewart, Luke M. and Hollerith, Simon and Zhou, Hengyun and Gullans, Michael J. and Yelin, Susanne F. and Greiner, Markus and Vuletić, Vladan and Cain, Madelyn and Lukin, Mikhail D.},
  journal={Nature},
  pages={39–46},
  year={2025},
  volume={649},
number={8095},
  publisher={Nature Publishing Group UK London},
doi={10.1038/s41586-025-09848-5},
url={https://doi.org/10.1038/s41586-025-09848-5}
}

@article{Browaeys2020,
	title={Many-body physics with individually controlled {Rydberg} atoms},
	author={Browaeys, Antoine and Lahaye, Thierry},
	journal={Nature Physics},
	volume={16},
	pages={132-142},
	year={2020},
	publisher={Nature Publishing Group},
	doi={10.1038/s41567-019-0733-z},
	url={https://doi.org/10.1038/s41567-019-0733-z}       
}

@misc{Cain2026,
      title={Shor's algorithm is possible with as few as 10,000 reconfigurable atomic qubits}, 
      author={Madelyn Cain and Qian Xu and Robbie King and Lewis R. B. Picard and Harry Levine and Manuel Endres and John Preskill and Hsin-Yuan Huang and Dolev Bluvstein},
      year={2026},
      eprint={2603.28627},
      archivePrefix={arXiv},
      url={https://arxiv.org/abs/2603.28627}, 
}

@article{Chiu2025,
	title={Continuous operation of a coherent 3,000-qubit system},
      author={Neng-Chun Chiu and Elias C. Trapp and Jinen Guo and Mohamed H. Abobeih and Luke M. Stewart and Simon Hollerith and Pavel Stroganov and Marcin Kalinowski and Alexandra A. Geim and Simon J. Evered and Sophie H. Li and Lisa M. Peters and Dolev Bluvstein and Tout T. Wang and Markus Greiner and Vladan Vuletić and Mikhail D. Lukin},
	journal={Nature},
	volume={646},
	number={},
	pages={1075–1080},
	year={2025},
	publisher={Nature Publishing Group},
	doi={10.1038/s41586-025-09596-6},
	url={https://doi.org/10.1038/s41586-025-09596-6}              
}

@article{Cao2024,
  title={Multi-qubit gates and {S}chr{\"o}dinger cat states in an optical clock},
  author={Cao, Alec and Eckner, William J. and Lukin Yelin, Theodor and Young, Aaron W. and Jandura, Sven and Yan, Lingfeng and Kim, Kyungtae and Pupillo, Guido and Ye, Jun and Darkwah Oppong, Nelson and Kaufman, Adam M.},
  journal={Nature},
  volume={634},
  number={8033},
  pages={315--320},
  year={2024},
  publisher={Nature Publishing Group UK London},
doi={10.1038/s41586-024-07913-z},
url={https://doi.org/10.1038/s41586-024-07913-z}
}

@article{Cooper2018,
  title = {Alkaline-Earth Atoms in Optical Tweezers},
  author = {Cooper, Alexandre and Covey, Jacob P. and Madjarov, Ivaylo S. and Porsev, Sergey G. and Safronova, Marianna S. and Endres, Manuel},
  journal = {Phys. Rev. X},
  volume = {8},
  issue = {4},
  pages = {041055},
  numpages = {19},
  year = {2018},
  month = {Dec},
  publisher = {American Physical Society},
  doi = {10.1103/PhysRevX.8.041055},
  url = {https://link.aps.org/doi/10.1103/PhysRevX.8.041055}
}

@article{Cornish2024,
  title={Quantum computation and quantum simulation with ultracold molecules},
  author={Cornish, Simon L and Tarbutt, Michael R and Hazzard, Kaden RA},
  journal={Nature Physics},
volume={20},
  pages={730–740},
  year={2024},
  publisher={Nature Publishing Group UK London},
url={https://doi.org/10.1038/s41567-024-02453-9} , 
doi={10.1038/s41567-024-02453-9}
}

@article{Defenu2023,
  title = {Long-range interacting quantum systems},
  author = {Defenu, Nicol\`o and Donner, Tobias and Macr\`{\i}, Tommaso and Pagano, Guido and Ruffo, Stefano and Trombettoni, Andrea},
  journal = {Rev. Mod. Phys.},
  volume = {95},
  issue = {3},
  pages = {035002},
  numpages = {70},
  year = {2023},
  month = {Aug},
  publisher = {American Physical Society},
  doi = {10.1103/RevModPhys.95.035002},
  url = {https://link.aps.org/doi/10.1103/RevModPhys.95.035002}
}

@misc{Desantis2026,
      title={Realization of a cavity-coupled {Rydberg} array}, 
      author={Jacopo De Santis and Balázs Dura-Kovács and Mehmet Öncü and Adrien Bouscal and Dimitrios Vasileiadis and Johannes Zeiher},
      year={2026},
      eprint={2602.12152},
      archivePrefix={arXiv},
      url={https://arxiv.org/abs/2602.12152},
      doi={10.48550/arXiv.2602.12152}
}

@article{Dumke2002,
	title = {Micro-optical Realization of Arrays of Selectively Addressable Dipole Traps: A Scalable Configuration for Quantum Computation with Atomic Qubits},
	author = {Dumke, R. and Volk, M. and M\"uther, T. and Buchkremer, F. B. J. and Birkl, G. and Ertmer, W.},
	journal = {Phys. Rev. Lett.},
	volume = {89},
	issue = {9},
	pages = {097903},
	numpages = {4},
	year = {2002},
	month = {Aug},
	publisher = {American Physical Society},
	doi = {10.1103/PhysRevLett.89.097903}
}

@article{Endres2016,
	author = {Endres, Manuel and Bernien, Hannes and Keesling, Alexander and Levine, Harry and Anschuetz, Eric R. and Krajenbrink, Alexandre and Senko, Crystal and Vuletic, Vladan and Greiner, Markus and Lukin, Mikhail D.},
	title = {Atom-by-atom assembly of defect-free one-dimensional cold atom arrays},
	volume = {354},
	number = {6315},
	pages = {1024--1027},
	year = {2016},
	doi = {10.1126/science.aah3752},
	publisher = {American Association for the Advancement of Science},
	issn = {0036-8075},
	URL = {http://science.sciencemag.org/content/354/6315/1024},
	journal = {Science}
}

@article{Evered2025,
  title={Probing the {K}itaev honeycomb model on a neutral-atom quantum computer},
  author={Evered, Simon J. and Kalinowski, Marcin and Geim, Alexandra A. and Manovitz, Tom and Bluvstein, Dolev and Li, Sophie H. and Maskara, Nishad and Zhou, Hengyun and Ebadi, Sepehr and Xu, Muqing and Campo, Joseph and Cain, Madelyn and Ostermann, Stefan and Yelin, Susanne F. and Sachdev, Subir and Greiner, Markus and Vuletić, Vladan and Lukin, Mikhail D.},
  journal={Nature},
  volume={645},
  number={8080},
  pages={341--347},
  year={2025},
  publisher={Nature Publishing Group UK London},
doi={10.1038/s41586-025-09475-0},
url={https://doi.org/10.1038/s41586-025-09475-0}
}

@article{Fang2025,
author = {Chengyu Fang  and Jared Miles  and Jonathan Goldwin  and Martin Lichtman  and Matthew Gillette  and Michael Bergdolt  and Sanket Deshpande  and Sam A. Norrell  and Preston Huft  and Mikhail A. Kats  and Mark Saffman },
title = {Interleaved dual-species arrays of single atoms using a passive optical element and one trapping laser},
journal = {Science Advances},
volume = {11},
number = {29},
pages = {eadw4166},
year = {2025},
doi = {10.1126/sciadv.adw4166},
URL = {https://www.science.org/doi/abs/10.1126/sciadv.adw4166}
}

@article{Finkelstein2024,
  title={Universal quantum operations and ancilla-based read-out for tweezer clocks},
  author={Finkelstein, Ran and Tsai, Richard Bing-Shiun and Sun, Xiangkai and Scholl, Pascal and Direkci, Su and Gefen, Tuvia and Choi, Joonhee and Shaw, Adam L and Endres, Manuel},
  journal={Nature},
  volume={634},
  number={8033},
  pages={321--327},
  year={2024},
  publisher={Nature Publishing Group UK London},
	doi={10.1038/s41586-024-08005-8},
	url={https://doi.org/10.1038/s41586-024-08005-8}        
}

@article{Graham2023,
author = {T. M. Graham and E. Oh and M. Saffman},
journal = {Appl. Opt.},
number = {12},
pages = {3242--3251},
publisher = {Optica Publishing Group},
title = {Multiscale architecture for fast optical addressing and control of large-scale qubit arrays},
volume = {62},
month = {Apr},
year = {2023},
url = {https://opg.optica.org/ao/abstract.cfm?URI=ao-62-12-3242},
doi = {10.1364/AO.484367}
}

@article{Grass2025,
  title = {Colloquium: Synthetic quantum matter in nonstandard geometries},
  author = {Grass, Tobias and Bercioux, Dario and Bhattacharya, Utso and Lewenstein, Maciej and Nguyen, Hai Son and Weitenberg, Christof},
  journal = {Rev. Mod. Phys.},
  volume = {97},
  issue = {1},
  pages = {011001},
  numpages = {31},
  year = {2025},
  month = {Mar},
  publisher = {American Physical Society},
  doi = {10.1103/RevModPhys.97.011001},
  url = {https://link.aps.org/doi/10.1103/RevModPhys.97.011001}
}

@article{Grinkemeyer2025,
author = {Brandon Grinkemeyer  and Elmer Guardado-Sanchez  and Ivana Dimitrova  and Danilo Shchepanovich  and G. Eirini Mandopoulou  and Johannes Borregaard  and Vladan Vuletić  and Mikhail D. Lukin },
title = {Error-detected quantum operations with neutral atoms mediated by an optical cavity},
journal = {Science},
volume = {387},
number = {6740},
pages = {1301-1305},
year = {2025},
doi = {10.1126/science.adr7075},
URL = {https://www.science.org/doi/abs/10.1126/science.adr7075},
}

@article{Gyger2024,
  title = {Continuous operation of large-scale atom arrays in optical lattices},
  author = {Gyger, Flavien and Ammenwerth, Maximilian and Tao, Renhao and Timme, Hendrik and Snigirev, Stepan and Bloch, Immanuel and Zeiher, Johannes},
  journal = {Phys. Rev. Res.},
  volume = {6},
  issue = {3},
  pages = {033104},
  numpages = {9},
  year = {2024},
  month = {Jul},
  publisher = {American Physical Society},
  doi = {10.1103/PhysRevResearch.6.033104},
  url = {https://link.aps.org/doi/10.1103/PhysRevResearch.6.033104}
}

@article{Gross2021,
  title={Quantum gas microscopy for single atom and spin detection},
  author={Gross, Christian and Bakr, Waseem S},
  journal={Nature Physics},
  volume={17},
  number={12},
  pages={1316--1323},
  year={2021},
  publisher={Nature Publishing Group UK London},
doi={10.1038/s41567-021-01370-5},
url={https://doi.org/10.1038/s41567-021-01370-5}
}

@article{Holman2026,
  title   = {Trapping of single atoms in metasurface optical tweezer arrays},
  author  = {Holman, Aaron and Xu, Yuan and Sun, Ximo and Wu, Jiahao and Wang, Mingxuan and Zhu, Zezheng and Seo, Bojeong and Yu, Nanfang and Will, Sebastian},
  journal = {Nature},
  volume={649},
  pages={859–865},
  year    = {2026},
  month   = {jan},
  day     = {14},
  doi     = {10.1038/s41586-025-09961-5}, 
  url     = {https://doi.org/10.1038/s41586-025-09961-5},
publisher={Nature Publishing Group UK London}
}

@article{Holzl2024,
  title = {Long-Lived Circular {Rydberg} Qubits of Alkaline-Earth Atoms in Optical Tweezers},
  author = {H\"olzl, C. and G\"otzelmann, A. and Pultinevicius, E. and Wirth, M. and Meinert, F.},
  journal = {Phys. Rev. X},
  volume = {14},
  issue = {2},
  pages = {021024},
  numpages = {11},
  year = {2024},
  month = {May},
  publisher = {American Physical Society},
  doi = {10.1103/PhysRevX.14.021024},
  url = {https://link.aps.org/doi/10.1103/PhysRevX.14.021024}
}

@article{Hsu2022,
  title = {Single-Atom Trapping in a Metasurface-Lens Optical Tweezer},
  author = {Hsu, T.-W. and Zhu, W. and Thiele, T. and Brown, M. O. and Papp, S. B. and Agrawal, A. and Regal, C. A.},
  journal = {PRX Quantum},
  volume = {3},
  issue = {3},
  pages = {030316},
  numpages = {11},
  year = {2022},
  month = {Aug},
  publisher = {American Physical Society},
  doi = {10.1103/PRXQuantum.3.030316},
  url = {https://link.aps.org/doi/10.1103/PRXQuantum.3.030316}
}

@article{Kaufman2021,
	title={Quantum science with optical tweezer arrays of ultracold atoms and molecules},
	author={Kaufman, Adam M and Ni, Kang-Kuen},
	journal={Nature Physics},
	volume={17},
	number={12},
	pages={1324--1333},
	year={2021},
	publisher={Nature Publishing Group},
	doi = {10.1038/s41567-021-01357-2}
}

@article{Kitson2025,
  title = {Rydberg atoms for electric field gradiometry},
  author = {Kitson, Philip and Chetcuti, Wayne J. and Birkl, Gerhard and Amico, Luigi and Polo, Juan},
  journal = {Phys. Rev. Res.},
  volume = {8},
  issue = {3},
  pages = {033010},
  numpages = {11},
  year = {2026},
  month = {Jul},
  publisher = {American Physical Society},
  doi = {10.1103/3jvx-hj7r},
  url = {https://link.aps.org/doi/10.1103/3jvx-hj7r}
}

@article{Lin2025,
  title = {{AI}-Enabled Parallel Assembly of Thousands of Defect-Free Neutral Atom Arrays},
  author = {Lin, Rui and Zhong, Han-Sen and Li, You and Zhao, Zhang-Rui and Zheng, Le-Tian and Hu, Tai-Ran and Wu, Hong-Ming and Wu, Zhan and Ma, Wei-Jie and Gao, Yan and Zhu, Yi-Kang and Su, Zhao-Feng and Ouyang, Wan-Li and Zhang, Yu-Chen and Rui, Jun and Chen, Ming-Cheng and Lu, Chao-Yang and Pan, Jian-Wei},
  journal = {Phys. Rev. Lett.},
  volume = {135},
  issue = {6},
  pages = {060602},
  numpages = {7},
  year = {2025},
  month = {Aug},
  publisher = {American Physical Society},
  doi = {10.1103/2ym8-vs82},
  url = {https://link.aps.org/doi/10.1103/2ym8-vs82}
}

@article{Langen2024,
  title={Quantum state manipulation and cooling of ultracold molecules},
  author={Langen, Tim and Valtolina, Giacomo and Wang, Dajun and Ye, Jun},
  journal={Nature Physics},
volume={20},
pages={702--712},
  year={2024},
  publisher={Nature Publishing Group UK London},
doi={10.1038/s41567-024-02423-1},
url={https://doi.org/10.1038/s41567-024-02423-1}
}

@misc{Li2025,
      title={Fast, continuous and coherent atom replacement in a neutral atom qubit array}, 
      author={Yiyi Li and Yicheng Bao and Michael Peper and Chenyuan Li and Jeff D. Thompson},
      year={2025},
      eprint={2506.15633},
      archivePrefix={arXiv},
      url={https://arxiv.org/abs/2506.15633}, 
}

@misc{Lim2026,
      title={Defect-free arrays at the thousand-atom scale in a {4-K} cryogenic environment}, 
      author={Desiree Lim and Hadriel Mamann and Grégoire Pichard and Lilian Bourachot and Arvid Lindberg and Clotilde Hamot and Hugo Le Bars and Florian Fasola and Siddhy Tan and Gwennolé Cournez and Sylvain Dutartre and Thierry Cartry and Sylvain Lemettre and Richard Hostein and Julien Paris and Franck Ferreyrol and Andréa Collardey and Adrien Signoles and Thierry Lahaye and Corentin Monmeyran and Bruno Ximenez},
      year={2026},
      eprint={2604.07205},
      archivePrefix={arXiv},
      url={https://arxiv.org/abs/2604.07205},
      doi={10.48550/arXiv.2604.07205}
}

@article{Mittenbuhler2025,
  title = {Model-based real-time synthesis of acousto-optically generated laser-beam patterns and tweezer arrays},
  author = {Mittenb\"uhler, Marcel and Sturm, Lukas and Schlosser, Malte and Birkl, Gerhard},
  journal = {Phys. Rev. Appl.},
  volume = {24},
  issue = {6},
  pages = {064046},
  numpages = {14},
  year = {2025},
  month = {Dec},
  publisher = {American Physical Society},
  doi = {10.1103/d3tx-3tg8},
  url = {https://link.aps.org/doi/10.1103/d3tx-3tg8}
}

@article{Manetsch2025,
	title={A tweezer array with 6100 highly coherent atomic qubits},
	author={Hannah J. Manetsch and Gyohei Nomura and Elie Bataille and Kon H. Leung and Xudong Lv and Manuel Endres},
	journal={Nature},
	volume={647},
	number={},
	pages={60–67},
	year={2025},
	doi={10.1038/s41586-025-09641-4},
url={https://doi.org/10.1038/s41586-025-09641-4},
	publisher={Nature Publishing Group}
}

@misc{Menssen2026,
      title={Strategic Plan for Neutral Atom Quantum Computation}, 
      author={Adrian J. Menssen and Tout Wang and Michael Gullans and Tom Manovitz and Jacob M. Taylor and Jason Cong and Josiah Sinclair and Ziv Aqua and Daniel J. Blumenthal and J. Pablo Bonilla Ataides and Johannes Borregaard and Antoine Browaeys and Paola Cappellaro and Soonwon Choi and Alexandre Cooper and Robin Côté and Jacob P. Covey and Alexandre Dauphin and Ivana Dimitrova and Matt Eichenfield and Dirk Englund and Jacob Freedman and Akihisa Goban and Brandon Grinkemeyer and Andi Gu and Ruonan Han and Dominik Hangleiter and Aram W. Harrow and Liang Jiang and Eun-ah Kim and Felix W. Knollmann and Aleksander Kubica and Thierry Lahaye and Lucas Lassabliere and Joonho Lee and Bingzhao Li and Mo Li and Wan-Hsuan Lin and Mikhail D. Lukin and Varun Menon and Thomas Propson and Akbar Safari and Mark Saffman and Pascal Scholl and Alexander Schuckert and Giulia Semeghini and Jonathan Simon and David Spierings and Daniel Bochen Tan and Shai Tsesses and Vladan Vuletic and Hanrui Wang and Hanyu Wang and Susanne Yelin and Johannes Zeiher and Hengyun Zhou},
      year={2026},
      eprint={2607.21554},
      archivePrefix={arXiv},
      url={https://arxiv.org/abs/2607.21554}, 
}

@article{Morgado2021,
	author = {Morgado,M.  and Whitlock,S. },
	title = {Quantum simulation and computing with {Rydberg}-interacting qubits},
	journal = {AVS Quantum Science},
	volume = {3},
	number = {2},
	pages = {023501},
	year = {2021},
	doi = {10.1116/5.0036562},
	URL = {https://doi.org/10.1116/5.0036562}
}

@article{Nakamura2024,
  title = {Hybrid Atom Tweezer Array of Nuclear Spin and Optical Clock Qubits},
  author = {Nakamura, Yuma and Kusano, Toshi and Yokoyama, Rei and Saito, Keito and Higashi, Koichiro and Ozawa, Naoya and Takano, Tetsushi and Takasu, Yosuke and Takahashi, Yoshiro},
  journal = {Phys. Rev. X},
  volume = {14},
  issue = {4},
  pages = {041062},
  numpages = {14},
  year = {2024},
  month = {Dec},
  publisher = {American Physical Society},
  doi = {10.1103/PhysRevX.14.041062},
  url = {https://link.aps.org/doi/10.1103/PhysRevX.14.041062}
}

@article{Norcia2024,
  title = {Iterative Assembly of ${}^{171}$$\mathrm{Yb}$ Atom Arrays with Cavity-Enhanced Optical Lattices},
  author = {Norcia, M. A. and Kim, H. and Cairncross, W. B. and Stone, M. and Ryou, A. and Jaffe, M. and Brown, M. O. and Barnes, K. and Battaglino, P. and Bohdanowicz, T. C. and Brown, A. and Cassella, K. and Chen, C.-A. and Coxe, R. and Crow, D. and Epstein, J. and Griger, C. and Halperin, E. and Hummel, F. and Jones, A. M. W. and Kindem, J. M. and King, J. and Kotru, K. and Lauigan, J. and Li, M. and Lu, M. and Megidish, E. and Marjanovic, J. and McDonald, M. and Mittiga, T. and Muniz, J. A. and Narayanaswami, S. and Nishiguchi, C. and Paule, T. and Pawlak, K. A. and Peng, L. S. and Pudenz, K. L. and Rodr\'{\i}guez P\'erez, D. and Smull, A. and Stack, D. and Urbanek, M. and van de Veerdonk, R. J. M. and Vendeiro, Z. and Wadleigh, L. and Wilkason, T. and Wu, T.-Y. and Xie, X. and Zalys-Geller, E. and Zhang, X. and Bloom, B. J.},
  journal = {PRX Quantum},
  volume = {5},
  issue = {3},
  pages = {030316},
  numpages = {13},
  year = {2024},
  month = {Jul},
  publisher = {American Physical Society},
  doi = {10.1103/PRXQuantum.5.030316},
  url = {https://link.aps.org/doi/10.1103/PRXQuantum.5.030316}
}

@article{Pause2023,
  title = {Reservoir-based deterministic loading of single-atom tweezer arrays},
  author = {Pause, Lars and Preuschoff, Tilman and Sch\"affner, Dominik and Schlosser, Malte and Birkl, Gerhard},
  journal = {Phys. Rev. Res.},
  volume = {5},
  issue = {3},
  pages = {L032009},
  numpages = {6},
  year = {2023},
  month = {Jul},
  publisher = {American Physical Society},
  doi = {10.1103/PhysRevResearch.5.L032009},
  url = {https://link.aps.org/doi/10.1103/PhysRevResearch.5.L032009}
}

@article{Pause2024,
author = {Lars Pause and Lukas Sturm and Marcel Mittenb\"{u}hler and Stephan Amann and Tilman Preuschoff and Dominik Sch\"{a}ffner and Malte Schlosser and Gerhard Birkl},
journal = {Optica},
number = {2},
pages = {222--226},
publisher = {Optica Publishing Group},
title = {Supercharged two-dimensional tweezer array with more than 1000 atomic qubits},
volume = {11},
month = {Feb},
year = {2024},
url = {https://opg.optica.org/optica/abstract.cfm?URI=optica-11-2-222},
doi = {10.1364/OPTICA.513551}
}

@article{Pichard2024,
  title = {Rearrangement of individual atoms in a 2000-site optical-tweezer array at cryogenic temperatures},
  author = {Pichard, Gr\'egoire and Lim, Desiree and Bloch, \'Etienne and Vaneecloo, Julien and Bourachot, Lilian and Both, Gert-Jan and M\'eriaux, Guillaume and Dutartre, Sylvain and Hostein, Richard and Paris, Julien and Ximenez, Bruno and Signoles, Adrien and Browaeys, Antoine and Lahaye, Thierry and Dreon, Davide},
  journal = {Phys. Rev. Appl.},
  volume = {22},
  issue = {2},
  pages = {024073},
  numpages = {7},
  year = {2024},
  month = {Aug},
  publisher = {American Physical Society},
  doi = {10.1103/PhysRevApplied.22.024073},
  url = {https://link.aps.org/doi/10.1103/PhysRevApplied.22.024073}
}

@article{Qiao2025,
  title={Realization of a doped quantum antiferromagnet in a {Rydberg} tweezer array},
  author={Qiao, Mu and Emperauger, Gabriel and Chen, Cheng and Homeier, Lukas and Hollerith, Simon and Bornet, Guillaume and Martin, Romain and Gély, Bastien and Klein, Lukas and Barredo, Daniel and Geier, Sebastian and Chiu, Neng-Chun and Grusdt, Fabian and Bohrdt, Annabelle and Lahaye, Thierry and Browaeys, Antoine},
  journal={Nature},
  pages={1--7},
  year={2025},
  publisher={Nature Publishing Group UK London},
doi={10.1038/s41586-025-09377-1},
url={https://doi.org/10.1038/s41586-025-09377-1}
}

@article{Radnaev2024,
  title = {Universal Neutral-Atom Quantum Computer with Individual Optical Addressing and Nondestructive Readout},
  author = {Radnaev, A.G. and Chung, W.C. and Cole, D.C. and Mason, D. and Ballance, T.G. and Bedalov, M.J. and Belknap, D.A. and Berman, M.R. and Blakely, M. and Bloomfield, I.L. and Buttler, P.D. and Campbell, C. and Chopinaud, A. and Copenhaver, E. and Dawes, M.K. and Eubanks, S.Y. and Friss, A.J. and Garcia, D.M. and Gilbert, J. and Gillette, M. and Goiporia, P. and Gokhale, P. and Goldwin, J. and Goodwin, D. and Graham, T.M. and Guttormsson, C.J. and Hickman, G.T. and Hurtley, L. and Iliev, M. and Jones, E.B. and Jones, R.A. and Kuper, K.W. and Lewis, T.B. and Lichtman, M.T. and Majdeteimouri, F. and Mason, J.J. and McMaster, J.K. and Miles, J.A. and Mitchell, P.T. and Murphree, J.D. and Neff-Mallon, N.A. and Oh, T. and Omole, V. and Parlo Simon, C. and Pederson, N. and Perlin, M.A. and Reiter, A. and Rines, R. and Romlow, P. and Scott, A.M. and Stiefvater, D. and Tanner, J.R. and Tucker, A.K. and Vinogradov, I.V. and Warter, M.L. and Yeo, M. and Saffman, M. and Noel, T.W.},
  journal = {PRX Quantum},
  volume = {6},
  issue = {3},
  pages = {030334},
  numpages = {20},
  year = {2025},
  month = {Aug},
  publisher = {American Physical Society},
  doi = {10.1103/66s8-jj18},
  url = {https://link.aps.org/doi/10.1103/66s8-jj18}
}

@misc{Reichardt2024,
      title={Fault-tolerant quantum computation with a neutral atom processor}, 
      author={Ben W. Reichardt and Adam Paetznick and David Aasen and Ivan Basov and Juan M. Bello-Rivas and Parsa Bonderson and Rui Chao and Wim van Dam and Matthew B. Hastings and Andres Paz and Marcus P. da Silva and Aarthi Sundaram and Krysta M. Svore and Alexander Vaschillo and Zhenghan Wang and Matt Zanner and William B. Cairncross and Cheng-An Chen and Daniel Crow and Hyosub Kim and Jonathan M. Kindem and Jonathan King and Michael McDonald and Matthew A. Norcia and Albert Ryou and Mark Stone and Laura Wadleigh and Katrina Barnes and Peter Battaglino and Thomas C. Bohdanowicz and Graham Booth and Andrew Brown and Mark O. Brown and Kayleigh Cassella and Robin Coxe and Jeffrey M. Epstein and Max Feldkamp and Christopher Griger and Eli Halperin and Andre Heinz and Frederic Hummel and Matthew Jaffe and Antonia M. W. Jones and Eliot Kapit and Krish Kotru and Joseph Lauigan and Ming Li and Jan Marjanovic and Eli Megidish and Matthew Meredith and Ryan Morshead and Juan A. Muniz and Sandeep Narayanaswami and Ciro Nishiguchi and Timothy Paule and Kelly A. Pawlak and Kristen L. Pudenz and David Rodríguez Pérez and Jon Simon and Aaron Smull and Daniel Stack and Miroslav Urbanek and René J. M. van de Veerdonk and Zachary Vendeiro and Robert T. Weverka and Thomas Wilkason and Tsung-Yao Wu and Xin Xie and Evan Zalys-Geller and Xiaogang Zhang and Benjamin J. Bloom},
      year={2024},
      eprint={2411.11822},
      archivePrefix={arXiv},
      url={https://arxiv.org/abs/2411.11822}, 
}

@article{Schaffner2020,
author = {Dominik Sch\"{a}ffner and Tilman Preuschoff and Simon Ristok and Lukas Brozio and Malte Schlosser and Harald Giessen and Gerhard Birkl},
journal = {Opt. Express},
number = {6},
pages = {8640--8645},
publisher = {Optica Publishing Group},
title = {Arrays of individually controllable optical tweezers based on {3D}-printed microlens arrays},
volume = {28},
month = {Mar},
year = {2020},
url = {https://opg.optica.org/oe/abstract.cfm?URI=oe-28-6-8640},
doi = {10.1364/OE.386243},
}

@article{Schaffner2024,
  title = {Quantum Sensing in Tweezer Arrays: Optical Magnetometry on an Individual-Atom Sensor Grid},
  author = {Sch\"affner, Dominik and Schreiber, Tobias and Lenz, Fabian and Schlosser, Malte and Birkl, Gerhard},
  journal = {PRX Quantum},
  volume = {5},
  issue = {1},
  pages = {010311},
  numpages = {8},
  year = {2024},
  month = {Jan},
  publisher = {American Physical Society},
  doi = {10.1103/PRXQuantum.5.010311},
  url = {https://link.aps.org/doi/10.1103/PRXQuantum.5.010311}
}

@article{Schlosser2011,
	author = {Schlosser, Malte and Tichelmann, Sascha and Kruse, Jens and Birkl, Gerhard},
	title = {Scalable architecture for quantum information processing with atoms in optical micro-structures},
	journal = {Quantum Information Processing},
	publisher = {Springer Netherlands},
	pages = {907-924},
	volume = {10},
	issue = {6},
	year = {2011},
	doi={10.1007/s11128-011-0297-z}
}

@article{Schlosser2012,
	doi = {10.1088/1367-2630/14/12/123034},
	url = {https://doi.org/10.1088%2F1367-2630%2F14%2F12%2F123034},
	year = 2012,
	month = {dec},
	publisher = {{IOP} Publishing},
	volume = {14},
	number = {12},
	pages = {123034},
	author = {M Schlosser and J Kruse and C Gierl and S Teichmann and S Tichelmann and G Birkl},
	title = {Fast transport, atom sample splitting and single-atom qubit supply in two-dimensional arrays of optical microtraps},
	journal = {New Journal of Physics}
}

@article{Schlosser2023,
	title = {Scalable Multilayer Architecture of Assembled Single-Atom Qubit Arrays in a Three-Dimensional {Talbot} Tweezer Lattice},
	author = {Schlosser, Malte and Tichelmann, Sascha and Sch\"affner, Dominik and Ohl de Mello, Daniel and Hambach, Moritz and Sch\"utz, Jan and Birkl, Gerhard},
	journal = {Phys. Rev. Lett.},
	volume = {130},
	issue = {18},
	pages = {180601},
	numpages = {7},
	year = {2023},
	month = {May},
	publisher = {American Physical Society},
	doi = {10.1103/PhysRevLett.130.180601},
	url = {https://link.aps.org/doi/10.1103/PhysRevLett.130.180601}
}

@article{Schymik2022,
  title = {In situ equalization of single-atom loading in large-scale optical tweezer arrays},
  author = {Schymik, Kai-Niklas and Ximenez, Bruno and Bloch, Etienne and Dreon, Davide and Signoles, Adrien and Nogrette, Florence and Barredo, Daniel and Browaeys, Antoine and Lahaye, Thierry},
  journal = {Phys. Rev. A},
  volume = {106},
  issue = {2},
  pages = {022611},
  numpages = {5},
  year = {2022},
  month = {Aug},
  publisher = {American Physical Society},
  doi = {10.1103/PhysRevA.106.022611},
  url = {https://link.aps.org/doi/10.1103/PhysRevA.106.022611}
}

@article{Shaw2025,
  title={A cavity-array microscope for parallel single-atom interfacing},
  author={Shaw, Adam L and Soper, Anna and Shadmany, Danial and Kumar, Aishwarya and Palm, Lukas and Koh, Da-Yeon and Kaxiras, Vassilios and Taneja, Lavanya and Jaffe, Matt and Schuster, David I and others},
  journal={Nature},
  volume={650},
  pages={320–326},
  year={2026},
  publisher={Nature Publishing Group UK London},
  doi={10.1038/s41586-025-10035-9},
  url={https://doi.org/10.1038/s41586-025-10035-9}
}

@article{Halimeh2025,
  title={Cold-atom quantum simulators of gauge theories},
  author={Halimeh, Jad C and Aidelsburger, Monika and Grusdt, Fabian and Hauke, Philipp and Yang, Bing},
  journal={Nature Physics},
  volume={21},
  number={1},
  pages={25–36},
  year={2025},
  publisher={Nature Publishing Group UK London},
doi={10.1038/s41567-024-02721-8},
url={https://doi.org/10.1038/s41567-024-02721-8}
}

@article{Shen2025,
    author = {Shen, Heng and Zhang, Jing},
    title = {Entanglement-enhanced quantum metrology with neutral atom arrays},
    journal = {Natl. Sci. Rev.},
    volume = {12},
    number = {8},
pages = {nwaf149},
    year = {2025},
    month = {04},
    issn = {2095-5138},
    url = {https://doi.org/10.1093/nsr/nwaf149},
    doi = {10.1093/nsr/nwaf149},
}

@article{Tian2023,
	title = {Parallel Assembly of Arbitrary Defect-Free Atom Arrays with a Multitweezer Algorithm},
	author = {Tian, Weikun and Wee, Wen Jun and Qu, An and Lim, Billy Jun Ming and Datla, Prithvi Raj and Koh, Vanessa Pei Wen and Loh, Huanqian},
	journal = {Phys. Rev. Appl.},
	volume = {19},
	issue = {3},
	pages = {034048},
	numpages = {10},
	year = {2023},
	month = {Mar},
	publisher = {American Physical Society},
	doi = {10.1103/PhysRevApplied.19.034048},
	url = {https://link.aps.org/doi/10.1103/PhysRevApplied.19.034048}
}

@article{Wintersperger2023,
	title={Neutral atom quantum computing hardware: performance and end-user perspective},
	author={Wintersperger, Karen and Dommert, Florian and Ehmer, Thomas and Hoursanov, Andrey and Klepsch, Johannes and Mauerer, Wolfgang and Reuber, Georg and Strohm, Thomas and Yin, Ming and Luber, Sebastian},
	journal={EPJ Quantum Technology},
	volume={10},
	number={32},
pages={},
	year={2023},
	publisher={Springer},
url={https://doi.org/10.1140/epjqt/s40507-023-00190-1},
doi={10.1140/epjqt/s40507-023-00190-1},

}

@article{Yan2022,
	title = {Two-Dimensional Programmable Tweezer Arrays of Fermions},
	author = {Yan, Zoe Z. and Spar, Benjamin M. and Prichard, Max L. and Chi, Sungjae and Wei, Hao-Tian and Ibarra-Garc\'{\i}a-Padilla, Eduardo and Hazzard, Kaden R. A. and Bakr, Waseem S.},
	journal = {Phys. Rev. Lett.},
	volume = {129},
	issue = {12},
	pages = {123201},
	numpages = {6},
	year = {2022},
	month = {Sep},
	publisher = {American Physical Society},
	doi = {10.1103/PhysRevLett.129.123201},
	url = {https://link.aps.org/doi/10.1103/PhysRevLett.129.123201}
}

@article{ZhangBichen2024,
author = {Bichen Zhang and Pai Peng and Aditya Paul and Jeff D. Thompson},
journal = {Optica},
number = {2},
pages = {227--233},
publisher = {Optica Publishing Group},
title = {Scaled local gate controller for optically addressed qubits},
volume = {11},
month = {Feb},
year = {2024},
url = {https://opg.optica.org/optica/abstract.cfm?URI=optica-11-2-227},
doi = {10.1364/OPTICA.512155}
}

@article{Chew2024,
  title = {Ultraprecise holographic optical tweezer array},
  author = {Chew, Y. T. and Poitrinal, M. and Tomita, T. and Kitade, S. and Mauricio, J. and Ohmori, K. and de L\'es\'eleuc, S.},
  journal = {Phys. Rev. A},
  volume = {110},
  issue = {5},
  pages = {053518},
  numpages = {10},
  year = {2024},
  month = {Nov},
  publisher = {American Physical Society},
  doi = {10.1103/PhysRevA.110.053518},
  url = {https://link.aps.org/doi/10.1103/PhysRevA.110.053518}
}

@article{ZhangBichen2026,
  title={Logical qubits with erasure conversion using metastable neutral atoms},
author={Bichen Zhang and Genyue Liu and Guillaume Bornet and Sebastian P. Horvath and Pai Peng and Shuo Ma and Shilin Huang and Shruti Puri and Jeff D. Thompson},
journal={Nature Physics},
volume={22},
number={},
pages={910–916},
year={2026},
publisher={Nature Publishing Group},
url={https://doi.org/10.1038/s41567-026-03309-0},
doi={https://doi.org/10.1038/s41567-026-03309-0}
}

@misc{Zhao2026,
      title={Towards Ultra-High-Rate Quantum Error Correction with Reconfigurable Atom Arrays}, 
      author={Chen Zhao and Casey Duckering and Andi Gu and Nishad Maskara and Hengyun Zhou},
      year={2026},
      eprint={2604.16209},
      archivePrefix={arXiv},
      url={https://arxiv.org/abs/2604.16209}, 
}
\end{document}